\documentclass[sigplan,10pt,nonacm]{acmart}
\renewcommand\footnotetextcopyrightpermission[1]{}
\usepackage{booktabs}
\usepackage{graphicx}
\usepackage{multirow}
\usepackage{enumitem}
\usepackage{amsmath}
\usepackage{algorithm}
\usepackage[noend]{algorithmic}
\usepackage{hyperref}
\usepackage{xcolor}
\usepackage{tikz}
\usepackage{pgfplots}
\usepackage{array}
\usepackage{dblfloatfix}
\usetikzlibrary{shapes.geometric, arrows, positioning, calc}

\definecolor{plotred}{HTML}{E41A1C}
\definecolor{plotblue}{HTML}{377EB8}
\definecolor{plotgreen}{HTML}{4DAF4A}
\definecolor{plotpurple}{HTML}{8C88C5}
\definecolor{plotorange}{HTML}{E67E22}
\definecolor{plotteal}{HTML}{3CAF8C}

\newcommand{\method}{DeltaLog}
\newcolumntype{C}[1]{>{\centering\arraybackslash}m{#1}}

\title[\method{}]{\method{}: Deferred Materialization of Recurrent States for Linear Attention Decoding}

\author{Junqing Lin}
\email{linjunqing@mail.ustc.edu.cn}
\orcid{0009-0007-1455-8725}
\affiliation{
    \institution{University of Science and Technology of China}
    \city{Hefei}
    \country{China}
}
\author{Jingwei Sun}
\authornote{Co-corresponding author}
\email{sunjw@ustc.edu.cn}
\orcid{0000-0001-5098-1503}
\affiliation{
    \institution{University of Science and Technology of China}
    \city{Hefei}
    \country{China}
}

\author{Guangzhong Sun}
\orcid{0000-0002-0794-7681}
\email{gzsun@ustc.edu.cn}
\affiliation{
    \institution{University of Science and Technology of China}
    \city{Hefei}
    \country{China}
}

\makeatletter
\let\old@fnsymbol\@fnsymbol
\def\@fnsymbol#1{%
  \ifnum#1=1\relax
    \textdagger
  \else
    \old@fnsymbol{#1}%
  \fi
}
\makeatother

\begin{document}
\settopmatter{printfolios=true}
\begin{abstract}

Linear attention models eliminate the quadratic prefix computation and context-growing KV cache of softmax attention by replacing pairwise token interactions with recurrent state updates. However, existing decoding implementations often materialize and write back the full recurrent state after every generated token, making state maintenance a major source of memory traffic, especially for models with large states and many heads.
This paper presents DeltaLog, a recurrent-state decoding scheme that reduces this overhead without changing the model semantics. Specifically, DeltaLog represents the recurrent state as a dense base state together with a bounded log of recent compact updates. Most decode steps append only compact update factors to this log, while periodic merge steps fold the accumulated updates back into the dense base state. Thus, the model observes the same dense state as in eager decoding, but most full-state write-backs are replaced by lightweight append operations.
We implement DeltaLog for GDN, KDA, and RWKV6 and integrate it into a prototype serving stack. Across these models, DeltaLog accelerates the recurrent-state update kernel by up to $1.86\times$, reduces profiled recurrent-state write traffic by up to $7.83\times$, and achieves $1.05$--$1.20\times$ end-to-end serving speedups over dense recurrent baselines.

\end{abstract}


\maketitle

\section{Introduction}

Large language models (LLMs) have become central to many production services~\cite{aminabadi2022deepspeed, yu2022orca, kwon2023efficient, zhong2024distserve}. However, as deployments scale, inference efficiency is increasingly limited by the softmax attention mechanism in Transformer-based LLMs~\cite{vaswani2017attention}, in which each token attends to all preceding tokens in the prefix. This design introduces two major efficiency bottlenecks. First, full-sequence attention requires quadratic token interactions, causing computational cost to grow rapidly with sequence length. Second, autoregressive decoding relies on a key-value (KV) cache to store previously computed keys and values, which can impose substantial GPU memory pressure when serving large batches or long sequences. These limitations have motivated systems optimizations for standard attention, including IO-aware kernels such as FlashAttention~\cite{dao2022flashattention, dao2024flashattention2, shah2024flashattention3} and serving systems that improve KV-cache management~\cite{sheng2023flexgen, kwon2023efficient, zhong2024distserve}.

A complementary line of work modifies the attention operator itself. Rather than computing softmax attention between the current token and all preceding tokens, recurrent linear-attention models~\cite{retnet,rwkv6,kimi_linear,gated_deltanet} maintain a fixed-size recurrent state that summarizes the prefix. During decoding, each step reads the previous state, applies an operator-specific transformation such as decay or gating, and incorporates the current token into the state. This update is compact, since it can be represented by two vectors instead of a full matrix.
However, the state that must be carried to the next token is still a dense matrix.
As a result, although recurrent linear attention avoids the context-length growth of the KV cache, it does not remove the cost of maintaining decode state. The bottleneck shifts from storing many token-level KV entries to repeatedly updating a dense recurrent state.

\begin{figure}[b]
\centering
\resizebox{0.95\columnwidth}{!}{
\definecolor{figink}{HTML}{1F2937}
\definecolor{figmuted}{HTML}{64748B}
\definecolor{figpanel}{HTML}{F8FAFC}
\definecolor{figbluefill}{HTML}{E8F1FC}
\definecolor{figblueedge}{HTML}{2F66A7}
\definecolor{figgreenfill}{HTML}{E8F7EE}
\definecolor{figgreenedge}{HTML}{2F855A}
\definecolor{figrededge}{HTML}{B42318}
\begin{tikzpicture}[
    font=\sffamily\scriptsize,
    >=stealth,
    panel/.style={draw=figink!16, fill=figpanel, rounded corners=5pt, line width=0.6pt},
    subpanel/.style={draw=figink!12, fill=white, rounded corners=4pt, line width=0.55pt},
    box/.style={draw=figink!40, rounded corners=2.5pt, align=center, inner sep=3pt, line width=0.7pt},
    factor/.style={box, fill=figgreenfill, draw=figgreenedge, line width=0.8pt},
    heading/.style={font=\sffamily\scriptsize\bfseries, text=figink},
    note/.style={font=\sffamily\scriptsize, text=figmuted, align=center},
    amplify/.style={->, draw=figrededge, line width=1.25pt}
]

\draw[panel] (0,0) rectangle (8.20,3.82);
\draw[subpanel, fill=figgreenfill!35] (0.26,0.32) rectangle (3.38,3.52);
\draw[subpanel, fill=figbluefill!50] (4.78,0.32) rectangle (7.92,3.52);
\draw[figink!12, dashed] (4.10,0.40) -- (4.10,3.40);

\node[heading, text=figgreenedge] at (1.82,3.25)
    {COMPACT FACTORS};
\node[factor, minimum width=6mm, minimum height=18mm] (akey) at (0.90,2.18)
    {$a_t$};
\node[font=\sffamily\large, text=figmuted] at (1.40,2.18) {$\otimes$};
\node[factor, minimum width=16mm, minimum height=6mm] (bvalue) at (2.50,2.18)
    {$b_t^\top$};
\node[note, rotate=90] at (0.47,2.18) {$d_k$};
\node[note] at (2.50,1.58) {$d_v$};
\node[note, text=figgreenedge] at (1.82,1.02) {$a_t, b_t^\top$ is specified compactly};

\draw[amplify] (3.43,2.12) -- (4.70,2.12);
\node[note, text=figrededge, font=\sffamily\tiny\bfseries] at (3.86,1.72)
    {WRITE AMPLIFICATION};

\node[heading, text=figrededge] at (6.35,3.25)
    {EAGER DENSE COMMIT};
\draw[fill=figbluefill, draw=figblueedge, rounded corners=2pt, line width=0.85pt]
    (5.10,1.00) rectangle (7.60,2.68);
\foreach \x in {5.4125,5.725,6.0375,6.350,6.6625,6.975,7.2875}
    \draw[figblueedge!28, line width=0.35pt] (\x,1.00) -- (\x,2.68);
\foreach \y in {1.336,1.672,2.008,2.344}
    \draw[figblueedge!28, line width=0.35pt] (5.10,\y) -- (7.60,\y);
\node[font=\sffamily\bfseries, text=figblueedge] at (6.35,1.96)
    {$S_t$};
\node[note] at (6.35,1.52) {$[B,H,d_k,d_v]$};

\end{tikzpicture}
}
\caption{The state-update tax. A token contributes compact factors with
$O(BH(d_k+d_v))$ values, but eager materialization writes back the full
$[B,H,d_k,d_v]$ recurrent state, touching $O(BHd_kd_v)$ elements per token.}
\label{fig:state_update_tax}
\end{figure}
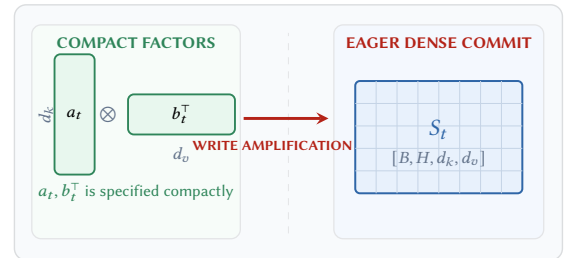

This recurrent update pattern creates a mismatch between the logical update required by linear-attention models and the physical memory traffic performed by existing implementations~\cite{yang2024fla, kwon2023efficient}. As shown in Figure~\ref{fig:state_update_tax}, for batch size $B$ and $H$ attention heads, the compact per-token update contains only $O(BH(d_k+d_v))$ values. In contrast, an eager implementation immediately materializes this update into the full recurrent state and writes back the dense $[B,H,d_k,d_v]$ tensor to high-bandwidth memory (HBM). This touches $O(BHd_kd_v)$ state elements at every generated token, even though the actual information added by the token is much smaller. We refer to this amplified per-token maintenance cost as the \emph{state-update tax}. This cost arises from the chosen materialization policy rather than from the recurrence itself.
This mismatch motivates a different execution strategy. Future decoding steps must observe the same logical state $S_t$, but $S_t$ does not need to be stored as a freshly materialized dense tensor after every token. Instead, recent compact updates can remain deferred, as long as subsequent state reads incorporate them exactly.

This paper proposes \method{}, a deferred materialization strategy for recurrent linear-attention decoding.
Rather than eagerly materializing this state as a fully updated dense tensor after every step, \method{} decomposes the state into a dense base state that stores older history and a bounded update log that stores recent compact updates. Each log entry stores the compact factors needed to reconstruct a recent state update. As a result, most decoding steps only append a compact log entry while leaving the dense base unchanged. To bound replay overhead, \method{} periodically merges the accumulated log entries into the dense base.
To implement \method{} efficiently on GPUs, we partition each head state along the value dimension. Each value tile stores the corresponding dense-base slice and value-side shard of the deferred update log, whereas the key-side component of each compact update is shared across value tiles and written only once per request–head. This layout exposes value-tile parallelism while avoiding redundant writes.
\method{} changes only the physical schedule by which state updates are committed, without modifying the model weights or logical recurrence. In exact arithmetic, it is algebraically equivalent to eager dense-state updates.

We evaluate \method{} with both decode kernels and a serving prototype. For one-token decoding across Gated DeltaNet (GDN)~\cite{gated_deltanet}, Kimi Delta Attention (KDA)~\cite{kimi_linear}, and RWKV6~\cite{rwkv6}, \method{} reduces latency relative to Fast Linear Attention (FLA)~\cite{yang2024fla}, a highly optimized dense recurrent-state baseline, by $1.19$--$1.69\times$ on H200 and $1.30$--$1.86\times$ on RTX 4090. In the serving prototype, the gains are smaller but positive, reaching $1.08$--$1.20\times$ on Qwen3.6-35B-A3B~\cite{qwen} and $1.05$--$1.12\times$ on Kimi-Linear-48B-A3B-Instruct~\cite{kimi_linear}. These results show that deferred recurrent-state write-back improves the targeted decode kernels and can translate into end-to-end serving benefits under the evaluated configurations.

This paper makes the following contributions.
\begin{itemize}[itemsep=10pt, topsep=10pt, parsep=0pt]
\item We identify and characterize the \emph{state-update tax}: the per-token overhead caused by eager dense recurrent-state materialization in large-batch recurrent linear attention serving.
\item We propose \method{}, a deferred-materialization execution strategy for recurrent linear-attention decoding. \method{} represents the state as a dense base augmented with a bounded log of compact updates, allowing most decoding steps to append compact factors rather than rewrite the full dense state.
\item We instantiate \method{} for one-token decoding in GDN, KDA, and RWKV6. Experiments show that our implementation preserves the original recurrence equations and improves decoding performance at the kernel level and in end-to-end prototype serving.
\end{itemize}

\begin{table}[b]
    \centering
    \caption{Representative recurrent formulations written in the components used by our systems analysis. The columns list the model-specific ingredients that instantiate Eq.~\ref{eq:unified_linear_attn}.}
    \label{tab:linear_attn_comparison}
    {\small
    \setlength{\tabcolsep}{4pt}
    \renewcommand{\arraystretch}{1.10}
    \begin{tabular}{@{} >{\raggedright\arraybackslash}m{0.20\linewidth} >{\centering\arraybackslash}m{0.20\linewidth} >{\centering\arraybackslash}m{0.36\linewidth} >{\centering\arraybackslash}m{0.12\linewidth} @{}}
        \toprule
        \textbf{Model} & \shortstack[c]{\textbf{State}\\\textbf{Transform}} & \shortstack[c]{\textbf{Update Pair}\\$(a_t,b_t)$} & \shortstack[c]{\textbf{Decay}\\\textbf{Form}} \\
        \midrule
        Linear Attention
            & $S_{t-1}$
            & $(\phi(k_t), v_t)$
            & none \\
        RetNet
            & $\gamma_t S_{t-1}$
            & $(k_t, v_t)$
            & scalar \\
        GLA
            & $D(\lambda_t) S_{t-1}$
            & $(k_t, v_t)$
            & diagonal \\
        RWKV6
            & $D(\lambda_t) S_{t-1}$
            & $(k_t, v_t)$
            & diagonal \\
        GDN
            & $D(\lambda_t) S_{t-1}$
            & $(k_t, \beta_t (v_t - k_t^\top S_{t-1}))$
            & scalar \\
        KDA
            & $D(\lambda_t) S_{t-1}$
            & $(k_t, \beta_t (v_t - k_t^\top S_{t-1}))$
            & diagonal \\
        \bottomrule
    \end{tabular}
    }
\end{table}

\section{Background and Motivation}

\subsection{Recurrent Form of Linear Attention}
\label{sec:background_linear_attention}

Softmax attention evaluates each query against the keys and values of preceding tokens, requiring the processed prefix to remain available as a token-level KV history. Linear attention instead uses a factorized similarity function that allows key--value contributions to be aggregated before a query arrives~\cite{katharopoulos2020transformers}. Ignoring optional normalization, causal linear attention can be written as the recurrence
\begin{equation}
S_t = S_{t-1} + \phi(k_t) v_t^\top,
\qquad
o_t = \phi(q_t)^\top S_t ,
\end{equation}
where $S_t$ aggregates the key--value interactions of the prefix through step $t$. Rather than retaining one key and value per preceding token, the recurrence carries a matrix state $S_t \in \mathbb{R}^{d_k \times d_v}$ whose shape is independent of context length.

Modern recurrent linear-attention models extend this basic additive recurrence with decay, gating, or delta-rule corrections~\cite{retnet,yang2024gla,rwkv6,gated_deltanet,kimi_linear}. For our systems analysis, these variants can be expressed by the same two operations: transforming the previous state and adding the outer product of two token-dependent vectors. Let
\begin{equation}
D(\ell) = \operatorname{diag}\!\left(\exp(\ell)\right).
\end{equation}
Absorbing any query feature map into $q_t$, a recurrent decoding step can then be written as
\begin{equation}
S_t = D(\lambda_t) S_{t-1} + a_t b_t^\top,
\qquad
o_t = \phi(q_t)^\top S_t ,
\label{eq:unified_linear_attn}
\end{equation}
where $\lambda_t \in \mathbb{R}^{d_k}$ is the per-step log-decay vector and $a_t \in \mathbb{R}^{d_k}$ and $b_t \in \mathbb{R}^{d_v}$ form the token update. Thus, the carried state is a dense $d_k \times d_v$ matrix, whereas each new token contributes an update represented by vectors of total size $O(d_k+d_v)$.
Table~\ref{tab:linear_attn_comparison} shows how representative models instantiate this common form. Linear attention~\cite{katharopoulos2020transformers} applies no decay, RetNet~\cite{retnet} uses scalar decay, and GLA~\cite{yang2024gla}, RWKV6~\cite{rwkv6}, and KDA~\cite{kimi_linear} use diagonal decay with different update semantics. Scalar decay is represented by sharing one value across the coordinates of $\lambda_t$, whereas diagonal decay uses a key-wise vector. Delta-rule models, including DeltaNet~\cite{yang2024parallelizing} and GDN~\cite{gated_deltanet}, additionally construct the value-side factor using the previous state. Eq.~\ref{eq:unified_linear_attn} therefore provides the common decode-time interface used in the remainder of our analysis.

\subsection{Recurrent State in Autoregressive Decoding}

LLM inference typically comprises two phases: \emph{prefill} and \emph{decode}~\cite{kwon2023efficient,zhong2024distserve}. The prefill phase processes the input prompt in parallel and constructs the state required for subsequent generation. The decode phase then generates tokens autoregressively, with each step depending on the state produced by the previous step. As a result, decode latency is sensitive not only to per-token arithmetic cost, but also to how the recurrent state is stored, accessed, and updated across steps.

In recurrent linear attention, prefill compresses the prompt into an initial recurrent state. During decode, each active request maintains one state slot per recurrent layer, instead of a token-level KV history. At step $t$, the decoder reads $S_{t-1}$, computes the output and token-specific update factors, and produces the logical state $S_t$ needed by the next step. Each state slot has a fixed shape independent of context length, but a serving batch must maintain and access these slots across all active requests and attention heads.

\begin{figure}[t]
\centering
\resizebox{0.475\textwidth}{!}{%
\begin{tikzpicture}
\begin{axis}[
    ybar,
    width=0.98\columnwidth,
    height=4.15cm,
    bar width=9.5pt,
    enlarge x limits=0.20,
    symbolic x coords={64,128,256},
    xtick=data,
    ymin=0,
    ymax=55,
    ytick={0,10,20,30,40,50},
    ylabel={Latency share (\%)},
    xlabel={Serving batch size ($B$)},
    axis lines*=left,
    axis line style={black!55},
    tick style={black!55},
    ymajorgrids=true,
    grid style={dashed, gray!22},
    label style={font=\scriptsize},
    tick label style={font=\scriptsize},
    nodes near coords,
    nodes near coords style={
        font=\tiny,
        rotate=90,
        anchor=west,
        yshift=1pt,
        /pgf/number format/fixed,
        /pgf/number format/precision=1
    },
    legend style={
        at={(0.5,1.02)},
        anchor=south,
        legend columns=3,
        draw=none,
        fill=none,
        font=\scriptsize,
        column sep=4pt
    },
    legend image code/.code={
        \draw[#1, draw=none] (0cm,-0.07cm) rectangle (0.16cm,0.14cm);
    },
    cycle list={
        {fill=plotpurple!92, draw=plotpurple!92},
        {fill=plotorange!88, draw=plotorange!88},
        {fill=plotteal!90, draw=plotteal!90}
    }
]
\addplot coordinates {(64,25.2) (128,35.6) (256,44.6)};
\addplot coordinates {(64,24.5) (128,34.2) (256,42.3)};
\addplot coordinates {(64,23.7) (128,32.6) (256,38.7)};
\legend{1k context, 2k context, 4k context}
\end{axis}
\end{tikzpicture}
}
\caption{Recurrent-state materialization occupies a growing share of full-model decode latency.}
\label{fig:linear_ratio}
\end{figure}
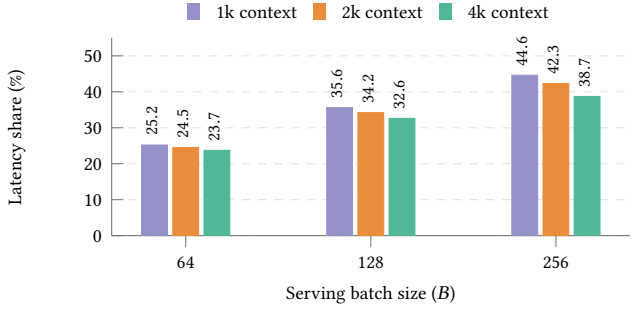

\subsection{The State-Update Tax in Serving}

Existing dense implementations typically materialize each state slot as a tensor and evaluate Eq.~\ref{eq:unified_linear_attn} eagerly~\cite{yang2024fla,zou2026kvbuffer}. Given the compact factors $a_t$ and $b_t$, they read the full state from HBM, apply the state transformation and outer-product update, and write the resulting dense state back to HBM. This design makes $S_t$ immediately available for the next decode step, but it places a full-state read--modify--write operation on the critical path of every generated token. Thus, replacing a context-growing KV cache with a fixed-shape recurrent state changes the structure of decode-state traffic, but does not necessarily make that traffic inexpensive.
The recurrence exposes an important asymmetry between the logical update and its eager physical realization. A new token contributes factors of total size $O(d_k+d_v)$, but immediately folding those factors into the state reads and writes $O(d_kd_v)$ elements. We call the memory traffic caused by this eager dense materialization the \emph{state-update tax}. The term separates the cost of physically maintaining the recurrent state from the compact information introduced by the token itself.

This tax scales with serving concurrency. For a batch of $B$ requests and $H$ heads, the runtime maintains a dense recurrent-state tensor of shape $[B,H,d_k,d_v]$, while the new token factors occupy only $O(BH(d_k+d_v))$ elements. Increasing the batch exposes more parallel decode work, but also increases the dense state traffic incurred at every iteration. A fixed per-request state can therefore become a large aggregate bandwidth cost.

To determine whether this cost remains visible beyond an isolated recurrent kernel, we measure its share of full-model decode latency in Qwen3.6-35B-A3B. Figure~\ref{fig:linear_ratio} shows a consistent concurrency trend across context lengths: for a 1k-token context, recurrent-state materialization grows from 25.2\% of decode latency at $B=64$ to 44.6\% at $B=256$. The corresponding shares at 2k and 4k contexts also rise monotonically with batch size. Eager state maintenance can therefore become a first-order serving bottleneck even though the recurrent state is independent of context length.

\subsection{Why the State-Update Tax Is Memory-Bound}
\label{sec:background_memory_wall}

The measurements above are consistent with a bandwidth bottleneck. Eager materialization converts a compact logical update into a large physical memory transaction: for every generated token, the runtime reads the dense $d_k \times d_v$ state from HBM, performs the state transition and compact update, and writes the entire result back. The amount of arithmetic and the amount of memory traffic both scale as $O(d_kd_v)$, but the update performs only a small constant amount of computation per state element.

The roofline model~\cite{williams2009roofline} characterizes this imbalance via arithmetic intensity, defined as computation per byte moved. Retaining only the dominant dense-state read and write traffic yields
\[
\text{AI}_{\text{dense}}
=
\frac{4 d_k d_v \ \text{FLOPs}}
{2 \cdot 4 d_k d_v \ \text{bytes}}
=
0.5 \ \text{FLOPs/byte}
\]
for a 32-bit state and approximately $1.0$ FLOPs/byte for a 16-bit state. This estimate deliberately excludes metadata traffic, output writes, vector loads, and model-specific additional arithmetic; it isolates the inherent cost of eagerly reading and writing the dense state.

\begin{figure}[h]
    \centering
    \includegraphics[width=0.45\textwidth]{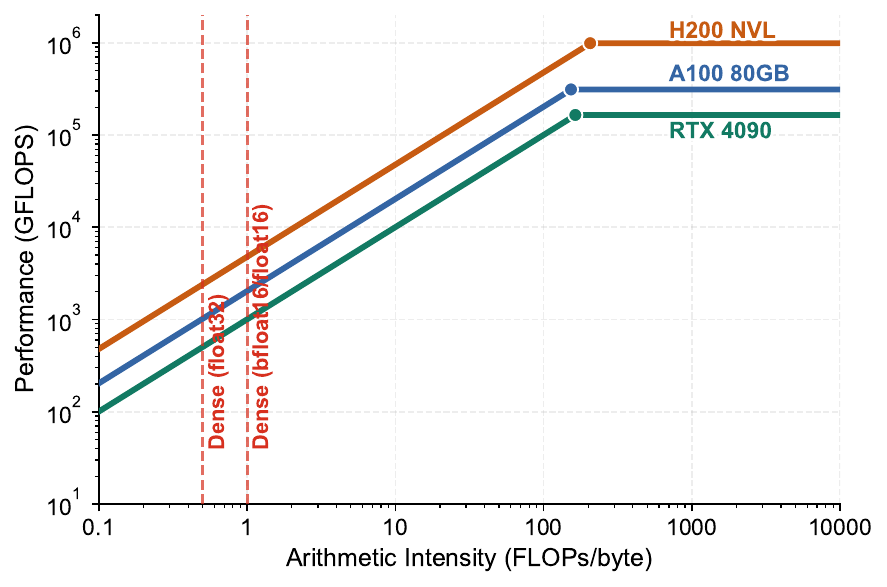}
    \caption{Roofline view of eager dense recurrent-state updates. Their arithmetic intensity is far below machine balance, making them strongly memory-bound.}
    \label{fig:roofline}
\end{figure}

\begin{figure*}[t]
    \centering
    \resizebox{0.9\textwidth}{!}{%
        \includegraphics[width=\textwidth]{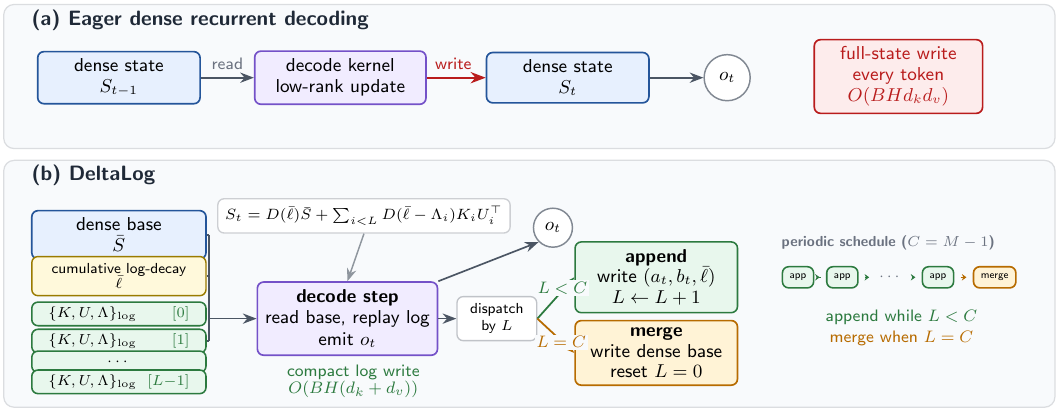}
    }
    \caption{\method{} overview. Eager recurrent decoding materializes the state at every token. \method{} instead stores a base plus a bounded compact update log, appends while $L < C$, and materializes the logical state into the dense base when $L=C$.}
    \label{fig:overview}
\end{figure*}

These arithmetic intensities are far below the machine balance of modern GPUs. For example, an H200-class accelerator provides on the order of $1{,}000$ TFLOPS of FP16/BF16 throughput and $4.8$ TB/s of HBM bandwidth, implying a balance of roughly $208$ FLOPs/byte. The gap of more than two orders of magnitude places eager dense-state updates deep in the memory-bound regime: compute is abundant, but per-token latency is dominated by HBM traffic. Figure~\ref{fig:roofline} visualizes this separation between dense recurrent updates and the machine-balance line.

The bottleneck becomes more severe as serving batch size grows. For batch size $B$ and $H$ attention heads, the recurrent state has shape $[B,H,d_k,d_v]$, so dense-state traffic scales linearly with the number of concurrently decoded sequences. With $B=512$, $H=32$, $d_k=d_v=128$, and a 32-bit state, the recurrent state occupies 1~GiB; one eager decode step moves roughly 2~GiB for the state read and write alone. This scaling explains the trend in Figure~\ref{fig:linear_ratio}: eager materialization creates a bandwidth-dominated latency floor even though the logical update introduced by each token remains compact. Avoiding this repeated dense write-back therefore requires a different physical representation of the recurrent state.

\section{Method}
\label{sec:method}

The preceding analysis indicates that the dominant inefficiency arises not from the recurrent update itself, but from eagerly materializing each update as dense HBM traffic. This observation motivates an alternative physical representation of the state for one-token recurrent decoding: one that preserves the original logical recurrence while avoiding dense per-token write-back. We therefore introduce \method{}, a deferred-materialization execution strategy for recurrent linear attention decoding. Figure~\ref{fig:overview} provides an overview of the execution workflow.

At a high level, \method{} represents the live recurrent state as a dense base augmented with a bounded log of recent compact updates, rather than overwriting the full dense state after every token. Let $M$ denote the merge interval. During the first $M-1$ steps of each cycle, new updates are appended to the log; at the \(M\)-th step, the accumulated logical state is materialized back into the dense base. Crucially, \method{} modifies only the \emph{physical} representation of the state: in exact arithmetic, the logical state observed by the model is identical to that produced by eager dense materialization. Model weights, operator equations, and prefill behavior are unchanged. The dense base is still read at every decoding step to preserve access to long-range history. Thus, the optimization specifically targets the write path, replacing dense per-token write-back with compact appends and one amortized dense write per cycle.

The remainder of this section develops the method from three progressively more concrete perspectives: the append--merge decoding cycle, the base-plus-log state invariant together with its traffic model, and the kernel- and operator-level realizations adopted in our implementation.

\subsection{Append--Merge Decode Cycle}
\label{sec:method_overview}
\label{sec:method_append_merge}

\method{} replaces eager dense write-back with an append--merge decoding cycle. During the common append step, the runtime records the current compact update in a bounded live log while leaving the dense recurrent state unchanged. At periodic merge steps, all deferred updates are rematerialized into the dense state, after which the log is cleared. By decoupling frequent compact updates from infrequent dense state materialization, this cycle avoids dense state write-back on most decoding steps while bounding the number of deferred updates that must be replayed. Algorithm~\ref{alg:deltalog_logical_step} summarizes the resulting logical decoding procedure.

Within each \method{} cycle, the live physical state consists of a dense base state $\bar{S}$ generated by the most recent materialization, a cumulative log-decay $\bar{\ell}$ applied since that materialization, and a bounded log of deferred compact updates.
Let $M$ denote the merge interval and $C=M-1$ the live-log capacity.
Each cycle comprises $C$ append steps followed by one merge step.
For a batch of requests, the physical state contains:
\begin{itemize}
    \item a dense FP32 base state $\bar{S} \in \mathbb{R}^{B \times H \times d_k \times d_v}$;
    \item a cumulative log-decay $\bar{\ell}$ in FP32, stored either as one value per head for head-wise scalar decay or as one value per key for per-key decay;
    \item bounded log buffers $K_{\log} \in \mathbb{R}^{B \times H \times C \times d_k}$ and $U_{\log} \in \mathbb{R}^{B \times H \times C \times d_v}$;
    \item one FP32 log-decay snapshot $\Lambda_{\log}$ per log entry, with the same scalar-or-vector structure as $\bar{\ell}$;
    \item the current live-log length $L \le C$.
\end{itemize}
Here $B$ is the serving batch size and $H$ is the number of heads.
The main structural difference across operators lies in the log-decay metadata: head-wise scalar decay requires $O(BH)$ metadata storage, whereas per-key decay requires $O(BHd_k)$.

\begin{algorithm}[t]
\caption{\method{} Logical Decode Step}
\label{alg:deltalog_logical_step}
\begin{algorithmic}[1]
\REQUIRE \parbox[t]{0.88\linewidth}{Base state $\bar{S}$, cumulative log-decay $\bar{\ell}$, live log $K_{\log}, U_{\log}, \Lambda_{\log}$, live-log length $L$, merge interval $M$, live-log capacity $C=M-1$, and current token tensors}
\STATE Evaluate any operator-defined readouts from the pre-update logical state represented by the dense base and live log.
\STATE Update cumulative log-decay: $\bar{\ell} \leftarrow \bar{\ell} + \lambda_t$.
\STATE Form the model-specific update factors $(a_t, b_t)$ and any post-update output required by the original recurrence.
\IF{$L < C$}
    \STATE Append $(a_t, b_t, \bar{\ell})$ into slot $L$ and set $L \leftarrow L + 1$.
\ELSE
    \STATE Materialize the represented logical state plus the current update into $\bar{S}$.
    \STATE Reset $\bar{\ell} \leftarrow 0$ and $L \leftarrow 0$.
\ENDIF
\STATE Return the step output.
\end{algorithmic}
\end{algorithm}

\paragraph{Append stage.}
When $L<C$, the runtime follows the append path.
It first evaluates any operator-defined readouts that must be taken from the pre-update logical state.
This ordering is required to preserve the original recurrence semantics: quantities defined on the old state must be evaluated before the transition, while quantities defined on the new state are evaluated afterward.
In the evaluated Delta-rule operators, the old-state correction is precisely such a pre-update readout.

The runtime then folds the current log-decay into the cumulative log-decay,
\begin{equation}
\bar{\ell} \leftarrow \bar{\ell} + \lambda_t .
\end{equation}
After forming the update factors $(a_t,b_t)$, it appends them to the current log slot $j=L$:
\begin{equation}
K_{\log}[j] \leftarrow a_t,
\qquad
U_{\log}[j] \leftarrow b_t,
\qquad
\Lambda_{\log}[j] \leftarrow \bar{\ell},
\label{eq:append_update}
\end{equation}
and increments $L$.
The dense base is left unchanged on this path.
Thus, the append path is computationally inexpensive, although each append increases the replay set by one entry.

\paragraph{Merge stage.}
The merge operation bounds live-log growth.
At the $M$-th step of a cycle, when the live log reaches its capacity $L=C$, the runtime materializes the represented logical state together with the current update into the dense base.
After applying the current log-decay and constructing the current update factors, it writes
\begin{equation}
\bar{S}
\leftarrow
D(\bar{\ell}) \bar{S}
+
\sum_{i=0}^{L-1}
D\!\left(\bar{\ell} - \Lambda_{\log}[i]\right)
K_{\log}[i] U_{\log}[i]^\top
+
a_t b_t^\top.
\label{eq:merge_update}
\end{equation}
The runtime then resets $\bar{\ell}\leftarrow 0$, clears the log, and sets $L\leftarrow 0$.
After this reset, the physical representation contains only the dense base.
Append and merge therefore serve complementary purposes: append avoids dense write-back in the common case, whereas merge incurs this cost once per cycle to keep replay bounded.

\subsection{State Invariant and Traffic Model}
\label{sec:method_state_traffic}
\label{sec:method_traffic}

We now formalize the logical state induced by the append--merge cycle and derive the corresponding recurrent-state traffic model. The key invariant is that, at every decoding step, the dense base state together with the deferred update log is logically equivalent to the state that would have been obtained under eager dense decoding. Suppose that there are currently \(L\) deferred entries, where \(0 \leq L \leq C\). The logical state represented by the physical state is then given by
\begin{equation}
    S_t =
    D(\bar{\ell}) \bar{S}
    +
    \sum_{i=0}^{L-1}
    D\!\left(\bar{\ell} - \Lambda_{\log}[i]\right)
    K_{\log}[i] U_{\log}[i]^\top.
    \label{eq:delta_log_invariant}
\end{equation}
The first term is the dense base decayed by the cumulative log-decay since the last materialization.
The second term replays all deferred compact updates, where the factor
$D\!\left(\bar{\ell} - \Lambda_{\log}[i]\right)$ accounts exactly for the decay accumulated after entry $i$ was inserted.

\paragraph{Closure under the recurrence.}
Eq.~\ref{eq:delta_log_invariant} is useful only if one recurrent step maps the representation back to the same form.
This follows directly from Eq.~\ref{eq:unified_linear_attn}.
Consider first an append step.
Applying $D(\lambda_t)$ increments the cumulative log-decay from $\bar{\ell}$ to $\bar{\ell}+\lambda_t$.
When $L<C$, the new compact contribution $a_t b_t^\top$ is stored as one additional deferred entry,
\[
(K_{\log}[L], U_{\log}[L], \Lambda_{\log}[L])
=
(a_t, b_t, \bar{\ell}+\lambda_t).
\]
Under this update, each previous log entry continues to be weighted by the decay accumulated after it was inserted, and the new entry has zero subsequent decay at insertion time.
Therefore, in exact arithmetic, the represented state after the append is
\(
D(\lambda_t)S_{t-1}+a_t b_t^\top,
\)
which matches the eager dense recurrence.
When $L=C$, the merge step writes Eq.~\ref{eq:merge_update} into the dense base and then resets $\bar{\ell}\leftarrow 0$ and $L\leftarrow 0$.
After the reset, Eq.~\ref{eq:delta_log_invariant} reduces to
\(
S_t = \bar{S}.
\)
Thus, the merge step also preserves the invariant.
The representation is therefore closed under both append and merge steps, establishing equivalence to eager dense decoding in exact arithmetic.

\paragraph{Traffic analysis.}
We next compare the recurrent-state traffic of eager dense decoding and \method{}.
To fairly compare the two representations, we separate traffic shared by both paths from state traffic specific to each representation.
Let $T_{\mathrm{common}}$ denote current-token inputs, such as the update factors and log-decay $\lambda_t$, that are read by both eager dense decoding and \method{}.
Let $p_S$ be the number of bytes per dense-state element.
The recurrent-state footprint of one layer is
\(
C_{\mathrm{state}} = p_SBHd_kd_v .
\)
An eager dense implementation reads and writes this state at every decoding step, yielding
\begin{equation}
T_{\mathrm{dense}}
\approx
T_{\mathrm{common}} + 2C_{\mathrm{state}} .
\end{equation}

For \method{}, let $p_K$ and $p_U$ denote the bytes per element of the deferred key- and value-side factors, respectively.
Let $d_\lambda$ denote the log-decay width, where $d_\lambda=1$ for head-wise scalar decay and $d_\lambda=d_k$ for per-key decay.
The persistent cumulative log-decay requires
\(
C_{\mathrm{decay}} = 4BHd_\lambda ,
\)
and each log entry, including its FP32 log-decay snapshot, requires
\begin{equation}
C_{\mathrm{entry}} =
BH\left(p_Kd_k+p_Ud_v+4d_\lambda\right).
\end{equation}
Over an $M$-step cycle, \method{} writes $M-1$ log entries and replays
\(
\sum_{L=0}^{M-1} L = \frac{M(M-1)}{2}
\)
entries.
Assuming the replay stream reads each live entry once, the idealized average per-step log traffic is
\begin{equation}
C_{\log}(M)
\approx
\left(\frac{M-1}{M}+\frac{M-1}{2}\right)C_{\mathrm{entry}},
\label{eq:log_traffic}
\end{equation}
where the two terms correspond to amortized append writes and replay reads, respectively.
The dominant per-step traffic for \method{} is therefore
\begin{equation}
T_{\mathrm{delta}}
\approx
T_{\mathrm{common}}
+
C_{\mathrm{state}}
+
\frac{C_{\mathrm{state}}}{M}
+
C_{\log}(M)
+
2C_{\mathrm{decay}} .
\label{eq:amortized_traffic}
\end{equation}
Under the same idealized single-stream assumption, the final term accounts for reading and writing the persistent cumulative log-decay once per step.
We omit lower-order length and index metadata, as well as transient scratch traffic.

Hence, \method{} is beneficial when the avoided dense write traffic,
\(
C_{\mathrm{state}}\left(1-\frac{1}{M}\right),
\)
exceeds the additional deferred-representation traffic,
\(
C_{\log}(M)+2C_{\mathrm{decay}}.
\)
Increasing $M$ reduces the amortized dense-state write cost but increases replay traffic, so the optimal merge interval is operator- and kernel-dependent.

\subsection{Tiled Kernel Decomposition}
\label{sec:method_kernel_realization}

\begin{figure*}[t]
    \centering
    \resizebox{0.99\textwidth}{!}{%
    \definecolor{ownink}{HTML}{1F2937}
    \definecolor{ownmuted}{HTML}{64748B}
    \definecolor{ownpanel}{HTML}{F8FAFC}
    \definecolor{owndatafill}{HTML}{EAF2FF}
    \definecolor{owndataedge}{HTML}{2F66A7}
    \definecolor{ownkernelfill}{HTML}{EAF7EF}
    \definecolor{ownkerneledge}{HTML}{2F855A}
    \definecolor{ownsharedfill}{HTML}{FFF3E0}
    \definecolor{ownsharededge}{HTML}{B7791F}
    \definecolor{ownscratchfill}{HTML}{F1F5F9}
    \definecolor{ownmergefill}{HTML}{F4F1FF}
    \definecolor{ownmergeedge}{HTML}{6750A4}
    \begin{tikzpicture}[
        font=\sffamily\scriptsize,
        >=stealth,
        line cap=round,
        line join=round,
        panel/.style={draw=ownink!14, rounded corners=6pt, fill=ownpanel, line width=0.6pt},
        zone/.style={draw=ownink!12, rounded corners=5pt, fill=white, line width=0.55pt},
        block/.style={draw=ownink!42, rounded corners=3pt, line width=0.75pt, align=center, inner sep=3.6pt},
        data/.style={block, fill=owndatafill, draw=owndataedge, minimum width=2.10cm, minimum height=0.60cm},
        shared/.style={block, fill=ownsharedfill, draw=ownsharededge, minimum width=2.50cm, minimum height=0.72cm},
        kernel/.style={block, fill=ownkernelfill, draw=ownkerneledge, minimum width=2.02cm, minimum height=0.66cm},
        scratch/.style={block, fill=ownscratchfill, draw=ownmuted, minimum width=2.02cm, minimum height=0.72cm},
        title/.style={font=\sffamily\bfseries\scriptsize, text=ownink},
        note/.style={font=\sffamily\scriptsize, text=ownmuted, align=center},
        stagebadge/.style={circle, draw=ownink!15, fill=white, inner sep=1.3pt, font=\sffamily\tiny\bfseries, text=ownmuted},
        arr/.style={->, line width=0.88pt, draw=ownink!68},
        dasharr/.style={->, line width=0.78pt, draw=ownmuted, dashed},
        mergebox/.style={block, dashed, draw=ownmergeedge, fill=ownmergefill, minimum width=4.70cm, minimum height=0.64cm}
    ]
        \draw[panel] (0.0,0.0) rectangle (18.90,6.25);
        \draw[zone, fill=ownkernelfill!28] (0.70,0.48) rectangle (5.72,5.35);
        \draw[zone, fill=owndatafill!26] (6.02,0.48) rectangle (13.45,5.35);
        \draw[zone, fill=ownsharedfill!35] (13.75,0.48) rectangle (18.20,5.35);

        \node[title, anchor=west] at (0.78,5.84) {Append--merge path for one slot--head pair};

        \node[stagebadge] at (1.00,5.02) {1};
        \node[title, anchor=west] at (1.22,5.02) {shared prep};
        \node[stagebadge] at (6.34,5.02) {2};
        \node[title, anchor=west] at (6.56,5.02) {tile decode};
        \node[note, anchor=west] at (6.56,4.66) {$N_V$ value-tile work units};
        \node[stagebadge] at (14.05,5.02) {3};
        \node[title, anchor=west] at (14.27,5.02) {single writer};

        \coordinate (sharedmid) at (1.78,3.16);
        \coordinate (prepaxis) at (4.10,3.16);
        \coordinate (commitaxis) at (16.05,3.18);

        \node[shared, minimum width=1.94cm, text width=1.58cm, minimum height=0.84cm] (sharedlog) at (sharedmid) {shared state\\$K_{\log}, \Lambda_{\log}$\\$\bar{\ell}, L$};
        \node[block, minimum width=1.78cm, fill=white, text width=1.42cm, minimum height=0.58cm] (token) at ($(prepaxis)+(0,1.08)$) {$q_t, k_t, \lambda_t$};
        \node[kernel, minimum width=2.02cm, text width=1.58cm, minimum height=0.78cm] (prep) at (prepaxis) {\textsc{ReplayPrep}};
        \node[scratch, minimum width=2.00cm, text width=1.46cm, minimum height=0.76cm] (scratchbox) at ($(prepaxis)+(0,-1.13)$) {FP32 replay\\coeffs};

        \node[data, text width=1.74cm] (in0) at (7.22,4.02) {$\bar{S}[:,\mathcal{V}_0]$\\$U_{\log}[:,\mathcal{V}_0]$};
        \node[kernel, text width=1.54cm] (dec0) at (9.62,4.02) {\textsc{Append}\\\textsc{Decode} $\mathcal{V}_0$};
        \node[data, text width=1.74cm] (out0) at (12.20,4.02) {$o_{\mathcal{V}_0}$\\$U_{\log}[L,\mathcal{V}_0]$};

        \node[data, text width=1.74cm] (inj) at (7.22,3.07) {$\bar{S}[:,\mathcal{V}_j]$\\$U_{\log}[:,\mathcal{V}_j]$};
        \node[kernel, text width=1.54cm] (decj) at (9.62,3.07) {\textsc{Append}\\\textsc{Decode} $\mathcal{V}_j$};
        \node[data, text width=1.74cm] (outj) at (12.20,3.07) {$o_{\mathcal{V}_j}$\\$U_{\log}[L,\mathcal{V}_j]$};

        \node[data, text width=1.74cm] (inlast) at (7.22,2.12) {$\bar{S}[:,\mathcal{V}_*]$\\$U_{\log}[:,\mathcal{V}_*]$};
        \node[kernel, text width=1.54cm] (declast) at (9.62,2.12) {\textsc{Append}\\\textsc{Decode} $\mathcal{V}_*$};
        \node[data, text width=1.74cm] (outlast) at (12.20,2.12) {$o_{\mathcal{V}_*}$\\$U_{\log}[L,\mathcal{V}_*]$};

        \node[mergebox, text width=4.32cm] (mergecallout) at (9.76,0.99)
            {merge phase ($L=C$): same tiled grid\\writes dense $\bar{S}[:,\mathcal{V}_j]$};

        \node[shared, minimum width=2.64cm, text width=2.08cm, minimum height=0.88cm] (owner) at ($(commitaxis)+(0,0.20)$) {append owner\\$K_{\log}[L], \Lambda_{\log}[L]$};
        \node[shared, minimum width=2.18cm, text width=1.60cm, minimum height=0.82cm] (finalize) at ($(commitaxis)+(0,-0.94)$) {\textsc{Finalize}\\$\bar{\ell}, L$};
        \node[note, text width=2.40cm] (ownernote) at ($(commitaxis)+(0,-1.70)$) {append increment or merge reset};

        \coordinate (coeffbus) at (6.06,3.07);

        \draw[arr] (token.south) -- (prep.north);
        \draw[arr] (sharedlog.east) -- (prep.west);
        \draw[arr] (prep.south) -- (scratchbox.north);

        \node[note, anchor=west, text width=1.30cm] at (6.20,1.44) {reused by all tiles};
        \draw[dasharr] (scratchbox.east) -- ++(0.48,0) |- (coeffbus);
        \draw[dasharr, -] (6.06,2.12) -- (6.06,4.02);
        \draw[dasharr] (6.06,4.02) -- (in0.west);
        \draw[dasharr] (6.06,3.07) -- (inj.west);
        \draw[dasharr] (6.06,2.12) -- (inlast.west);

        \draw[arr] (in0.east) -- (dec0.west);
        \draw[arr] (dec0.east) -- (out0.west);
        \draw[arr] (inj.east) -- (decj.west);
        \draw[arr] (decj.east) -- (outj.west);
        \draw[arr] (inlast.east) -- (declast.west);
        \draw[arr] (declast.east) -- (outlast.west);

        \draw[arr] (out0.east) -- ++(0.52,0) |- (owner.west);
        \draw[arr] (owner.south) -- (finalize.north);
        \draw[dasharr] (declast.south) -- ([xshift=-0.14cm]mergecallout.north);
        \draw[dasharr] (mergecallout.east) -- ++(1.46,0) |- (finalize.west);
    \end{tikzpicture}
    }
    \caption{The append–merge kernel ownership path comprises three stages: shared replay preparation, parallel append decode across value tiles, and single-writer metadata commit. The merge path uses the same structure but writes back dense-base tiles.}
    \label{fig:append_kernel_ownership}
\end{figure*}
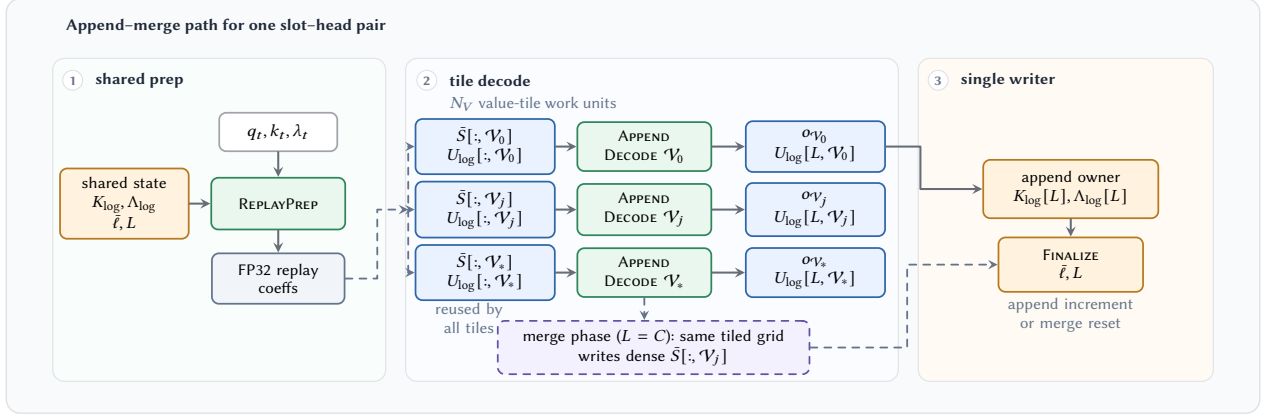

The append--merge cycle defines the logical procedure for decoding; however, an efficient implementation must also conform to the GPU execution model. Since each slot--head state contains a full \(d_k \times d_v\) matrix, assigning an entire state to a single work unit would severely limit parallelism and make it difficult to saturate memory bandwidth. We therefore partition the state along the value dimension. For each state slot and attention head, each work unit owns a dense-base tile of size \(d_k \times B_V\), where \(B_V\) is the tile width along the value dimension and \(\mathcal{V}_j\) denotes the value coordinates assigned to tile \(j\). This yields \(N_V=\lceil d_v/B_V\rceil\) value tiles, and the GPU execution grid is organized over state slots, attention heads, and value tiles.

This decomposition induces an asymmetry between value-local data and shared metadata. The dense-base tile and the corresponding shard of the value-side log, \(U_{\log}\), are local to each value tile; thus, each work unit can stream its assigned dense tile, update the corresponding output slice, and write to \(U_{\log}[L,\mathcal{V}_j]\) independently. In contrast, the key-side entry \(K_{\log}[L]\), the log-decay snapshot \(\Lambda_{\log}[L]\), the cumulative log-decay \(\bar{\ell}\), and the live-log length \(L\) are shared across all value tiles associated with the same slot and head.

This asymmetry creates two coordination requirements. First, replay coefficients should be computed once per slot--head pair rather than redundantly by every value tile. Replaying a previous entry requires key-side quantities such as \(q_t^\top K_{\log}[i]\) and, for Delta-rule operators, the correction projection \(k_t^\top K_{\log}[i]\), together with the row-wise rescaling induced by \(D(\bar{\ell}-\Lambda_{\log}[i])\). These terms depend on the current head and log entry but are independent of the value tile. Second, the shared append metadata must be updated by a single logical owner, while tile-local work units remain responsible only for their output slices and value-log shards.

We address both issues with a staged append decomposition based on ownership, illustrated in Figure~\ref{fig:append_kernel_ownership}. When replay reuse is profitable, \textsc{ReplayPrep} runs once for each state slot and head, computes the reusable key-side replay coefficients, and stores them in transient FP32 scratch space. \textsc{AppendDecode} then executes in parallel across value tiles: each work unit streams its dense-base tile, loads the corresponding tile-local $U_{\log}$ shard, reuses the prepared coefficients, computes its output slice, and appends the new $U_{\log}[L,\mathcal{V}_j]$ shard. A designated writer records the shared $K_{\log}[L]$ and $\Lambda_{\log}[L]$ entries, while \textsc{Finalize} commits the cumulative-log-decay update and increments $L$ exactly once.

Merge follows the same value-tiled execution structure, but differs in its commit target. Instead of appending a compact log entry, each tile-local work unit accumulates the replayed contributions associated with its dense-base tile and writes the fully materialized FP32 tile back to HBM. After this materialization step, a single metadata update resets both the cumulative log-decay and the live-log length.

\subsection{Operator Instantiations}
\label{sec:method_operator_instantiation}

We instantiate the base-plus-log design for three operators: GDN~\cite{gated_deltanet}, KDA~\cite{kimi_linear}, and RWKV6~\cite{rwkv6}. These operators all preserve the eager logical recurrence up to floating-point tolerance, but differ in two orthogonal implementation choices that govern replay cost in $C_{\log}(M)$: decay granularity and update semantics. Along these axes, GDN provides the simplest instantiation, with head-wise scalar decay and a Delta-rule correction; KDA increases the decay granularity to per-key vectors while retaining the correction path; and RWKV6 keeps the per-key log-decay metadata but removes the correction path by using raw-value updates.

GDN is a lightweight instantiation of the Delta rule. Its head-wise scalar decay requires only one FP32 cumulative log-decay per head. Although append operations still compute the Delta-rule old-state correction before committing each deferred entry, replay remains inexpensive because each deferred key-side factor is rescaled by a single scalar. Consequently, GDN benefits from the shared-\textsc{ReplayPrep} decomposition introduced in Section~\ref{sec:method_kernel_realization}: its coefficients can be reused across value tiles with minimal metadata overhead. The dense base state and log-decay metadata are stored in FP32. Deferred key-side factors $a_t$ use 16-bit storage, whereas deferred value-side factors $b_t$ are retained in FP32 to mitigate cancellation error along the correction path.

KDA retains GDN's Delta-rule corrected update but replaces head-wise scalar decay with per-key decay. Replay must therefore rescale each deferred key-side factor element-wise using the cumulative log-decay vector, which increases both arithmetic cost and metadata traffic. \textsc{ReplayPrep} still reuses key-side projections across value tiles, but its relative benefit is smaller because per-key scaling remains necessary. The dense base state, cumulative per-key log-decay vectors, and per-entry log-decay snapshots are stored in FP32. Deferred key-side factors $a_t$ use 16-bit storage, while deferred value-side factors $b_t$ remain in FP32 for the same cancellation-sensitive correction path as in GDN.

RWKV6 shares KDA's per-key log-decay metadata, including FP32 cumulative log-decay vectors and per-entry log-decay snapshots, but differs in update semantics. Because it uses the raw token value rather than a Delta-rule corrected update, replay only replays deferred key-side factors under per-key decay and avoids the old-state correction path. This simplification allows more replay work to be folded into the main decode kernel. The dense base state and log-decay metadata are stored in FP32, while deferred key- and value-side factors are stored in BF16.

\section{End-to-End System Implementation}
\label{sec:implementation}

This section describes how the proposed scheduling scheme is integrated into a graph-captured serving runtime. We prototype \method{} in a vLLM-based serving stack~\cite{kwon2023efficient}, replacing the eager dense decode path in the evaluated serving configurations while leaving model weights and prefill kernels unchanged. This integration raises three implementation challenges: replaying phase-specific decode graphs, managing the base-plus-log recurrent state, and preserving correctness with continuous batching.

\paragraph{Graph selection and execution.}
Because \method{} decomposes decoding into append and merge phases with distinct kernel schedules and state-update semantics, the full decode path cannot be represented by a single CUDA graph. Our prototype therefore captures two CUDA-graph variants during warmup for each uniform decode-batch descriptor: one for append and one for merge. These graphs expose the same model-level inputs and outputs but differ in how they commit recurrent state. The append graph records the current update in the bounded log, whereas the merge graph replays the logical state, writes it back to the dense base, and clears the deferred representation. During replay, the runtime tracks the phase associated with each captured uniform batch and selects the corresponding pre-captured graph, with the phase encoded in the graph-cache key. Thus, the additional host-side overhead is limited to phase selection and graph-cache lookup, rather than to launching the internal recurrent kernels individually.

\paragraph{State-slot lifecycle.}
To support the base-plus-log representation, we extend the recurrent-state allocator for each integrated linear-attention layer. In addition to the dense recurrent state produced during prefill, each allocated slot maintains bounded key- and value-log storage, cumulative log-decay metadata, per-entry log-decay snapshots, and live-log metadata. When a request is assigned a fresh slot, its prefill output initializes the dense base and the auxiliary representation is initialized as empty. When a slot is reused, only its auxiliary buffers and metadata are reset, leaving the states of other active requests unchanged.

\begin{figure*}[b]
    \centering
    \includegraphics[width=0.98\textwidth]{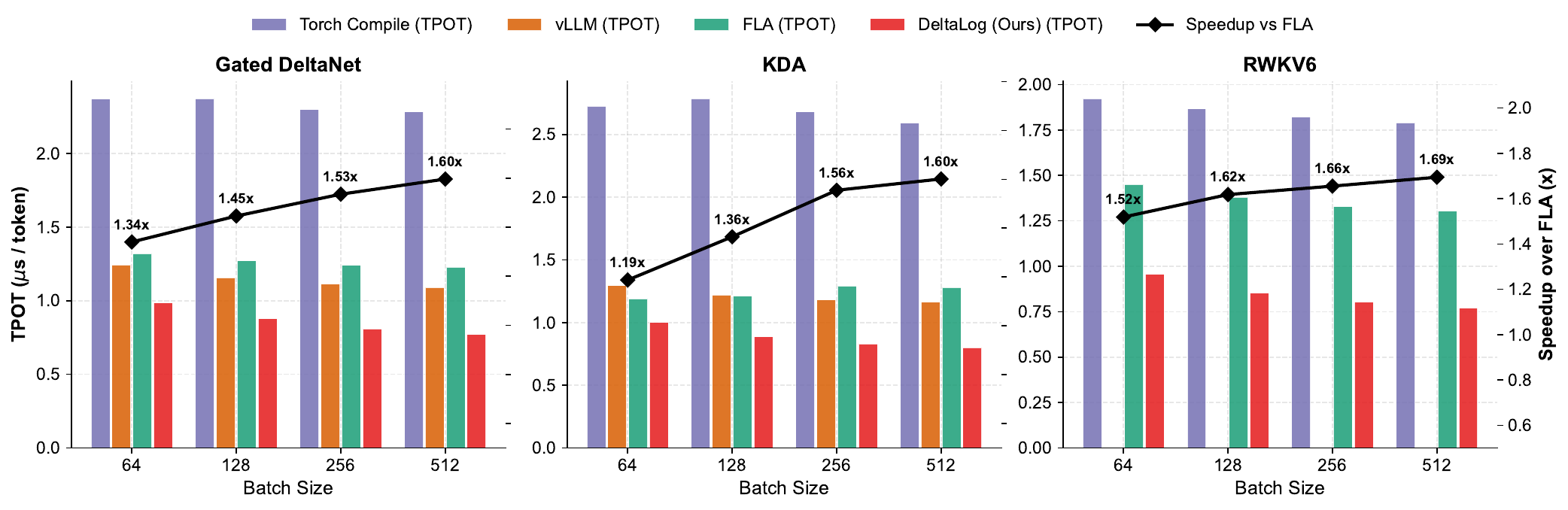}
    \caption{Single-token decode performance on H200. Bars show TPOT, and lines show \method{} speedup over FLA.}
    \label{fig:kernel_combined_h200}
\end{figure*}

\begin{figure*}[b]
    \centering
    \includegraphics[width=0.98\textwidth]{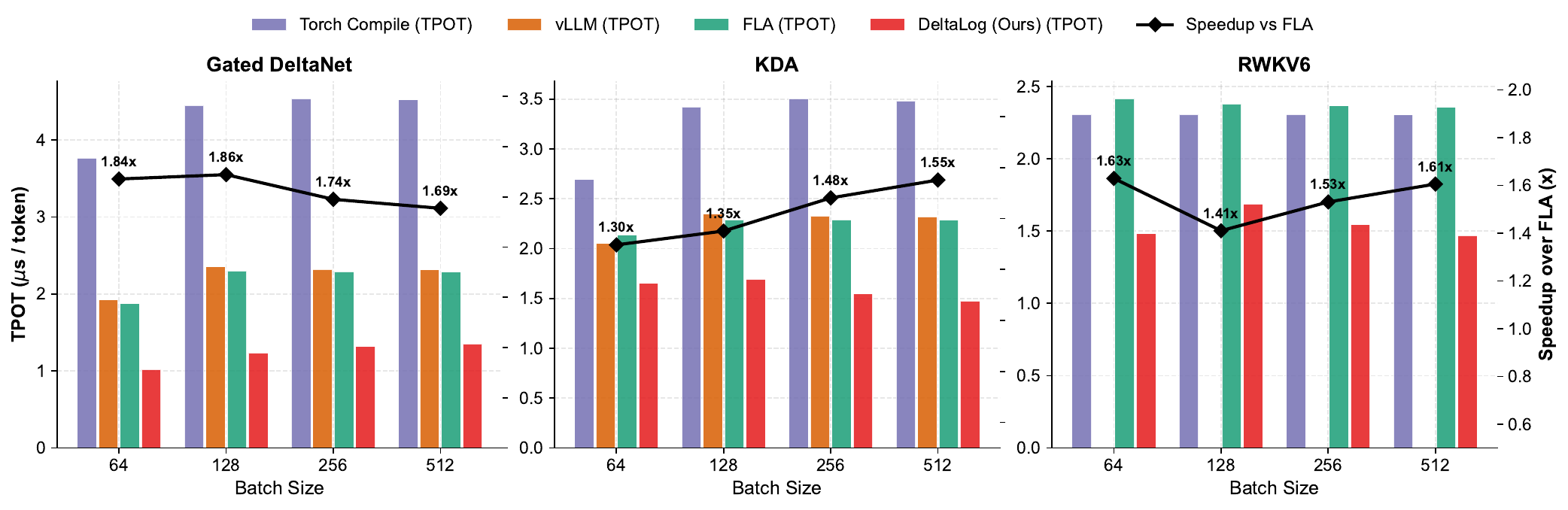}
    \caption{
    Single-token decode performance on RTX 4090. Bars show TPOT, and lines show \method{} speedup over FLA.
    }
    \label{fig:kernel_combined_4090}
\end{figure*}

\paragraph{Phase-specialized decode scheduling.}
Graph replay requires phase-aware batching rather than a single monolithic decode queue. The runtime therefore maintains separate append and merge queues: requests are normally decoded through append-graph replay, moved to the merge queue when they reach a merge boundary, and returned to the append queue after the corresponding merge graph materializes deferred updates into the dense base and clears the live log. This policy preserves continuous batching while satisfying the phase-specific graph interface. Newly admitted requests are initialized with an empty auxiliary representation and placed directly in the append queue, so they can batch with other append-phase requests without modifying existing slots and are never forced into merge executions for which they have no deferred updates.

\section{Experiments}
\label{sec:experiments}

We evaluate \method{} as a systematic optimization for recurrent decoding, focusing on three questions: whether deferred materialization improves decode performance, what mechanisms drive the improvement, and how much kernel-level benefit persists after integration into a serving stack. To this end, we present both kernel- and serving-level evidence. At the kernel level, we measure one-token decode latency across operators, batch sizes, and GPUs; use hardware counters to validate the write-back mechanism; analyze sensitivity to merge interval and recurrent-state size; and assess numerical behavior under mixed precision. At the serving level, we evaluate whether these gains persist in a batch-synchronous graph-mode prototype.

\subsection{Experimental Setup}
\label{sec:experimental_setup}

\paragraph{Hardware platforms.}
We evaluate kernel decode performance on two NVIDIA GPUs: an RTX 4090, representing a commodity single-GPU platform, and an H200 NVL, representing a data-center serving platform. End-to-end serving experiments are run on H200 only. Table~\ref{tab:gpu_platforms} summarizes the hardware context using vendor specifications. 

\begin{table}[h]
    \centering
    \caption{Hardware configuration.}
    \label{tab:gpu_platforms}
    {\small
    \setlength{\tabcolsep}{5pt}
    \renewcommand{\arraystretch}{1.15}
    \resizebox{0.98\linewidth}{!}{%
    \begin{tabular}{@{} c c c c c @{}}
        \toprule
        \multirow{2}{*}{\textbf{GPU}} &
        \textbf{Deployment} &
        \multirow{2}{*}{\textbf{Memory}} &
        \textbf{Peak Memory} &
        \textbf{Peak Tensor} \\
        &
        \textbf{Class} &
        &
        \textbf{Bandwidth} &
        \textbf{Throughput} \\
        \midrule
        RTX 4090 & commodity & 24 GB & 1008 GB/s & 165.2 TFLOPS \\
        H200 NVL & data-center & 141 GB & 4.8 TB/s & 1671 TFLOPS \\
        \bottomrule
    \end{tabular}%
    }
    }
\end{table}

\paragraph{Implementation scope.}
The experimental implementation is designed to separate kernel-level applicability from serving-level impact, while focusing exclusively on the decoding stage. We do not evaluate prefill, since \method{} targets recurrent-state updates during autoregressive token generation rather than the parallel computation used in prefill. For kernel-level evaluation, we implement \method{} as standalone Triton~\cite{tillet2019triton} decode kernels, enabling controlled comparisons across the recurrent operators considered in this study. For serving-level evaluation, we incorporate \method{} into the batch-synchronous vLLM-based prototype described in Section~\ref{sec:implementation}; this integration is used only for the full-model decoding experiments reported below.

\begin{table}[htbp]
    \centering
    \caption{Recurrent-kernel configurations. Shape reports $H/d_k/d_v$; log-decay metadata is FP32 for all operators.}
    \label{tab:operator_configs}
    {
    \setlength{\tabcolsep}{8pt}
    \renewcommand{\arraystretch}{1.12}
    \begin{tabular}{@{} l c c c c @{}}
        \toprule
        \textbf{Operator} & \textbf{Shape} & \textbf{State} & \shortstack[c]{\textbf{Log} $(K/U)$} & $M$ \\
        \midrule
        GDN   & 32/128/128 & FP32 & BF16/FP32 & 8 \\
        KDA   & 32/128/128 & FP32 & BF16/FP32 & 4 \\
        RWKV6 & 32/128/128 & FP32 & BF16/BF16 & 4 \\
        \bottomrule
    \end{tabular}
    }
\end{table}

\paragraph{Operators, models, and precision.}
Our kernel study covers GDN~\cite{gated_deltanet}, KDA~\cite{kimi_linear}, and RWKV6~\cite{rwkv6}. Table~\ref{tab:operator_configs} summarizes the standalone-kernel shapes, persistent-state and log precisions, and merge intervals selected from the sweep in Section~\ref{sec:sensitivity}. Activations are stored in \texttt{bfloat16}, whereas dense recurrent states and log-decay metadata are stored in \texttt{float32}, following the numerically stable dense configurations. GDN and KDA retain the value-side update log in FP32 because their Delta-rule corrections are more sensitive to cancellation; RWKV6 stores both log factors in BF16. The serving study evaluates the integrated GDN and KDA paths, using Qwen3.6-35B-A3B~\cite{qwen} for GDN and Kimi-Linear-48B-A3B-Instruct for KDA~\cite{kimi_linear}.

\paragraph{Baselines.}
For kernel experiments, we compare \method{} with dense recurrent baselines whenever available. \textbf{PyTorch Compile} is a dense PyTorch implementation optimized with \texttt{torch.compile}~\cite{ansel2024pytorch2}. \textbf{FLA} uses optimized Triton kernels from Fast Linear Attention~\cite{yang2024fla} and is the strongest dense baseline available for all three operator families in our local environment. \textbf{vLLM} denotes the dense recurrent kernel path used in our serving-stack integration~\cite{kwon2023efficient}; because it is available for GDN and KDA but not RWKV6, we restrict vLLM-relative claims to GDN and KDA.
For end-to-end serving, we use the same vLLM-based stack as the baseline, with dense recurrent decode enabled in place of \method{}. 

Baseline and \method{} runs use identical models, batch sizes, decode lengths, precision settings, and CUDA-graph replay configurations. Thus, differences in time per output token (TPOT) isolate the effect of replacing eager dense recurrent write-back with \method{} under an otherwise fixed serving configuration.

\subsection{Kernel Decode Performance}
\label{sec:kernel_perf}

We first isolate the recurrent decode kernel, where \method{} directly changes execution. A dense recurrent implementation reads and writes the full state at every generated token, whereas \method{} replaces most dense writes with compact log appends and amortizes materialization through periodic merges.
Figures~\ref{fig:kernel_combined_4090} and~\ref{fig:kernel_combined_h200} report one-token decode latency across operators and batch sizes. On H200, \method{} improves over FLA by $1.34$--$1.60\times$ for GDN, $1.19$--$1.60\times$ for KDA, and $1.52$--$1.69\times$ for RWKV6 over $B \in \{64,128,256,512\}$. On RTX 4090, the corresponding FLA-relative speedups are $1.69$--$1.86\times$ for GDN, $1.30$--$1.55\times$ for KDA, and $1.41$--$1.63\times$ for RWKV6. For H200 configurations with vLLM dense kernel results, \method{} improves over vLLM by $1.26$--$1.42\times$ for GDN and $1.29$--$1.46\times$ for KDA.

The main trend is that \method{} remains beneficial across both GPUs, although the gain is operator-dependent. GDN generally benefits the most because its head-wise scalar decay keeps the log metadata compact. KDA shares the same Delta-rule correction structure but uses per-key decay, which increases metadata and replay work; its speedup is therefore smaller at some shapes. RWKV6 also uses per-key decay, but its update path is simpler because it does not require the Delta-rule old-state correction, helping its speedups remain competitive despite the heavier metadata.

\subsection{Where the Speedup Comes From}
\label{sec:hbm_traffic}

We next examine the mechanism behind the speedup. We first profile one representative H200 GDN configuration to verify the write-back reduction directly. We then ablate the internal organization of the append path to measure the additional benefit of split replay across operators.

\begin{figure}[h]
    \centering
    \includegraphics[width=\linewidth]{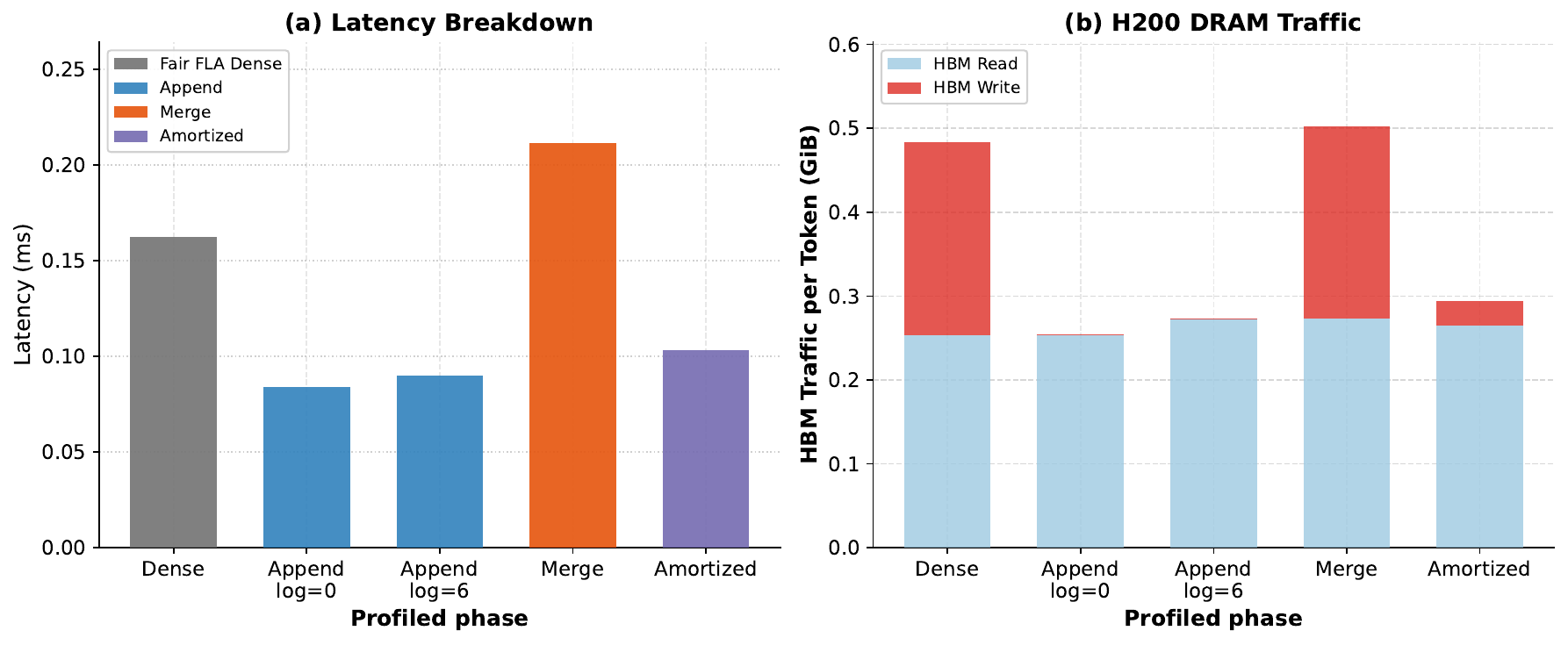}
    \caption{Latency and profiled DRAM traffic for GDN at $B=128$ and $M=8$. (a) reports dense decoding, append latency by log size, merge latency, and the amortized result. (b) reports the corresponding DRAM-traffic breakdown.}
    \label{fig:breakdown_analysis}
\end{figure}
\paragraph{Latency and Traffic Breakdown.}
\method{} is designed to reduce dense recurrent-state write-back. To test that mechanism directly, we profile one representative GDN configuration on H200 at $B=128$ with merge interval $M=8$. Figure~\ref{fig:breakdown_analysis} compares the dense reference with four \method{} views at the same shape: the first append step (live log size 0), the last append step before merge (live log size 6), the merge step that materializes a live log of size 7, and the resulting amortized per-token latency. The second panel reports the corresponding Nsight Compute DRAM counters.

Panel (a) shows the expected append--merge behavior. Append latency increases only marginally as the live log grows. The subsequent merge is more expensive because it materializes the dense state. However, this cost is incurred once per interval rather than once per token. Compared with the dense FLA reference latency at the same shape, all append steps remain below the dense baseline, while the merge cost is amortized over the full merge cycle.
Panel (b) provides consistent evidence from DRAM traffic. Dense GDN reads 0.253~GiB and writes 0.230~GiB, for total profiled DRAM traffic of 0.483~GiB. In \method{}, append steps read both the dense base and the live log, causing profiled read traffic to increase from 0.253~GiB at log size 0 to 0.273~GiB at log size 6. By contrast, append writes remain small because the append path commits only compact log entries and metadata, rather than materializing or persisting the full dense state. The merge restores the dense write-back cost, reading 0.274~GiB and writing 0.229~GiB. Amortized over $M=8$, \method{} writes only 0.0294~GiB per token, corresponding to a $7.83\times$ reduction in write traffic and a $1.64\times$ reduction in total profiled DRAM traffic relative to the dense baseline.

\begin{figure}[h]
    \centering
    \includegraphics[width=\linewidth]{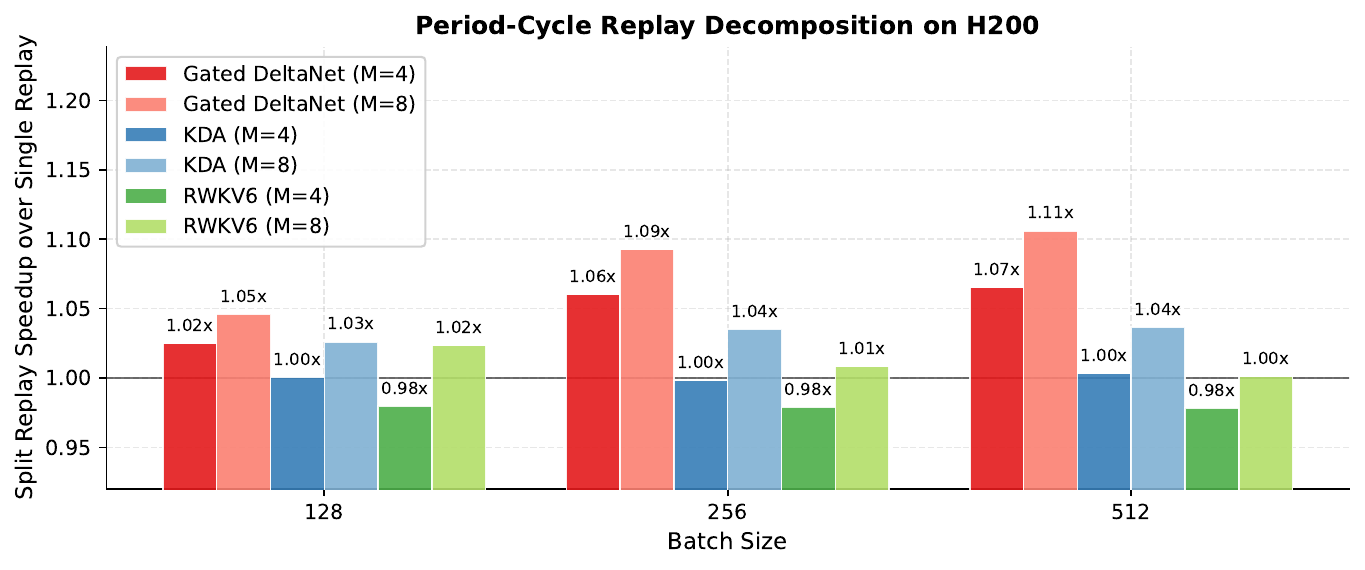}
    \caption{Split-replay ablation on H200 with CUDA Graph replay. Latencies are amortized over one merge cycle; GDN, KDA, and RWKV6 are shown at $M=4$ and $M=8$.}
    \label{fig:kernel_split_ablation}
\end{figure}

\paragraph{Split-Replay Ablation.}
We ablate the organization of the \method{} append path. Specifically, we compare a fused append kernel with a split path that precomputes replay dot products before append decode; both variants use the same metadata-finalization step. For fairness, we capture the full cycle of $M{-}1$ appends followed by one merge in a CUDA Graph and report amortized per-token latency.
Figure~\ref{fig:kernel_split_ablation} shows that split replay consistently improves GDN, yielding $1.02$--$1.07\times$ speedup at $M=4$ and $1.05$--$1.11\times$ at $M=8$ over $B \in \{128,256,512\}$. KDA exhibits smaller gains: it is nearly neutral at its selected $M=4$ configuration ($0.998$--$1.003\times$), but improves by $1.03$--$1.04\times$ at $M=8$. RWKV6 benefits the least, slowing down at $M=4$ ($0.979$--$0.980\times$) and reaching only $1.00$--$1.02\times$ at $M=8$.

These trends are explained by operator structure rather than total kernel cost alone. Split replay is beneficial only when duplicated live-log dot products constitute a substantial fraction of append work. This condition holds for GDN, particularly at $M=8$, because its decay metadata is head-wise scalar. By contrast, KDA and RWKV6 retain per-key decay operations after the split; KDA additionally applies the Delta-rule correction against the dense base state, while RWKV6 already has a relatively inexpensive fused replay path. Consequently, split replay is most effective when the eliminated replay computation outweighs the added prepare-kernel launch and scratch-memory overhead.

\subsection{Sensitivity to Merge Interval and State Size}
\label{sec:sensitivity}

\begin{figure}[t]
    \centering
    \includegraphics[width=\linewidth]{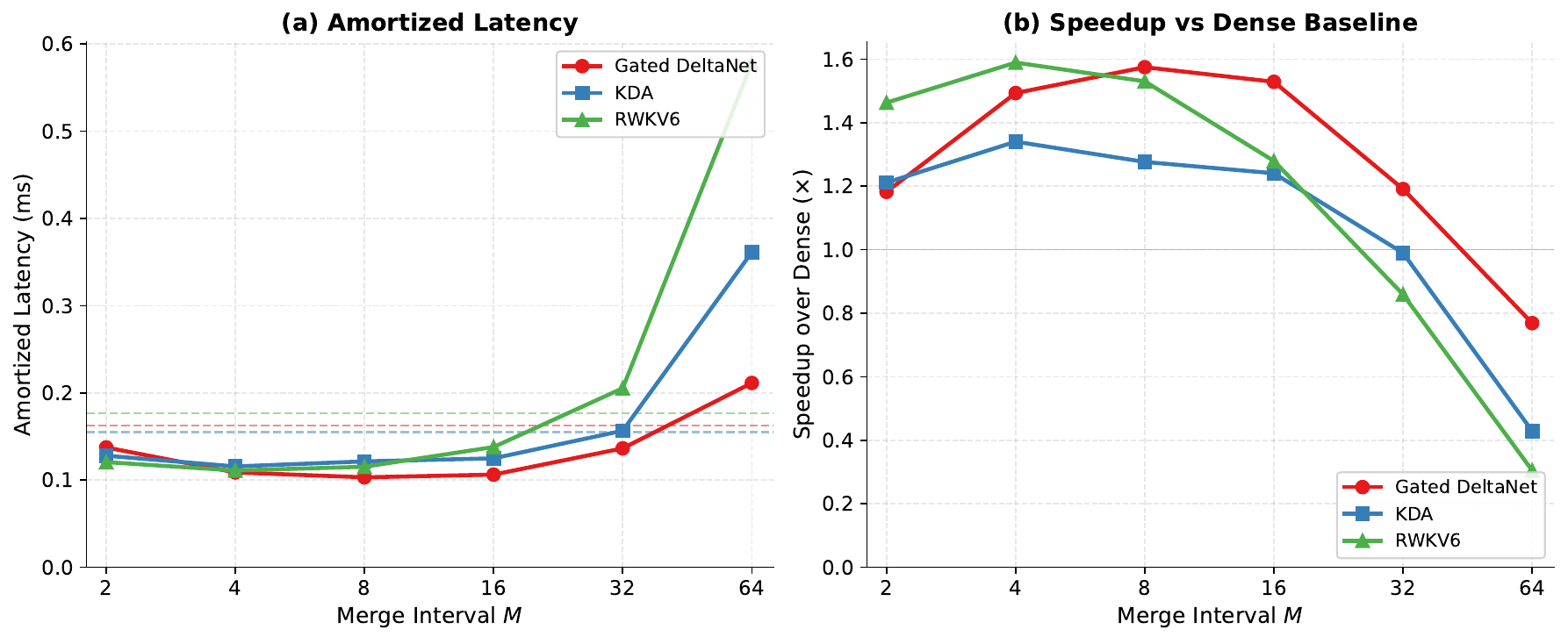}
    \caption{Interval sweep for \method{} at $B=128$. (a) Amortized per-token latency. (b) Speedup over FLA.}
    \label{fig:merge_interval_sweep}
\end{figure}

The merge interval $M$ controls the central trade-off in \method{}. A larger $M$ reduces the frequency of dense-state materialization, but it also increases live-log replay during append. Figure~\ref{fig:merge_interval_sweep} visualizes this trade-off on H200 at $B=128$ for $M \in \{2,4,8,16,32,64\}$, with nearby operator-specific candidates also included for selection.

The measured curves are U-shaped. Small intervals merge too frequently and recover only part of the avoided write-back cost. Very large intervals amortize merge more aggressively but increase append latency because the active log is longer. At $B=128$, the best FLA-relative intervals are GDN $M=8$ with 0.103~ms amortized latency and $1.57\times$ speedup, KDA $M=4$ with 0.116~ms and $1.34\times$, and RWKV6 $M=4$ with 0.110~ms and $1.60\times$.

\begin{figure}[h]
    \centering
    \includegraphics[width=0.75\linewidth]{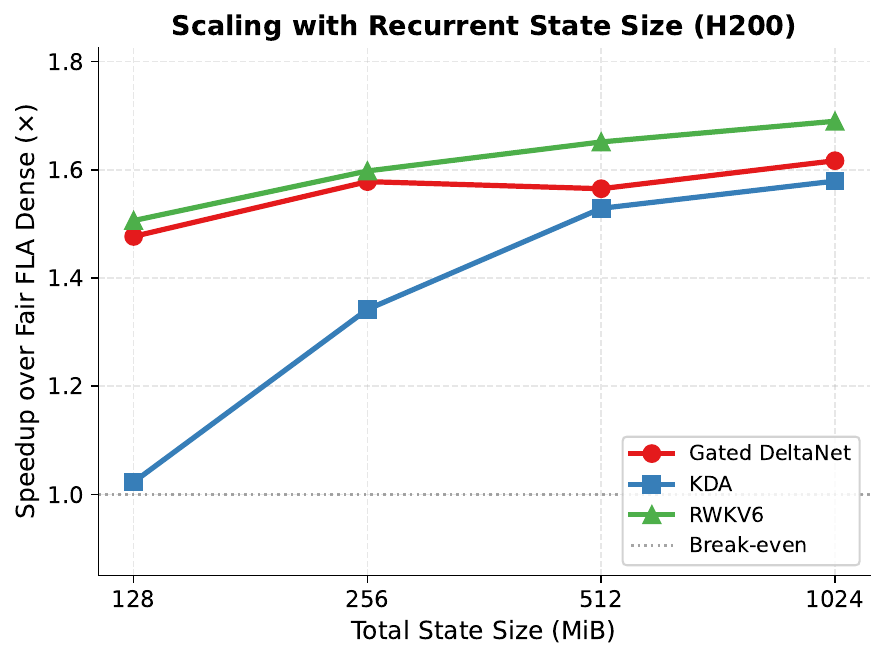}
    \caption{Scaling with total recurrent-state size on H200.}
    \label{fig:state_size_scaling}
\end{figure}

We also vary the total recurrent-state size from 128~MiB to 1024~MiB on H200 while keeping the operator dimensions fixed and changing batch size. Figure~\ref{fig:state_size_scaling} shows that \method{} becomes more effective as the recurrent state grows. Against FLA, GDN speedup increases from $1.48\times$ at 128~MiB to $1.62\times$ at 1024~MiB. KDA is close to parity at the smallest state size, with $1.02\times$ speedup at 128~MiB, but reaches $1.58\times$ at 1024~MiB. RWKV6 improves from $1.51\times$ to $1.69\times$ over the same range.
These results identify the main operating regime of the method. \method{} is most effective when recurrent-state write-back is a first-order cost. When the state is small, replay and metadata overhead can offset the saved write traffic. When the state is large, the avoided dense writes dominate the additional log work.

\subsection{Numerical Behavior}
\label{sec:numerical_correctness}

\method{} changes the order and timing of recurrent-state materialization. Although it is algebraically equivalent to eager dense decoding under exact arithmetic, the implemented kernels use mixed precision and a different accumulation order, which may introduce floating-point discrepancies. We therefore assess numerical behavior against dense references using 128-token kernel rollouts for $B \in \{64,128,256\}$.

All three operators satisfy both the repository tolerances and the BF16-aware benchmark tolerances in the selected-$M$ campaign. The maximum output relative errors are 0.00303 for GDN, 0.00450 for KDA, and 0.00376 for RWKV6; the corresponding maximum state relative errors are $7.92{\times}10^{-7}$, $4.61{\times}10^{-4}$, and $2.86{\times}10^{-7}$. These results support describing the kernels as tolerance-equivalent to dense references.

\begin{figure}[htbp]
    \centering
    \includegraphics[width=\linewidth]{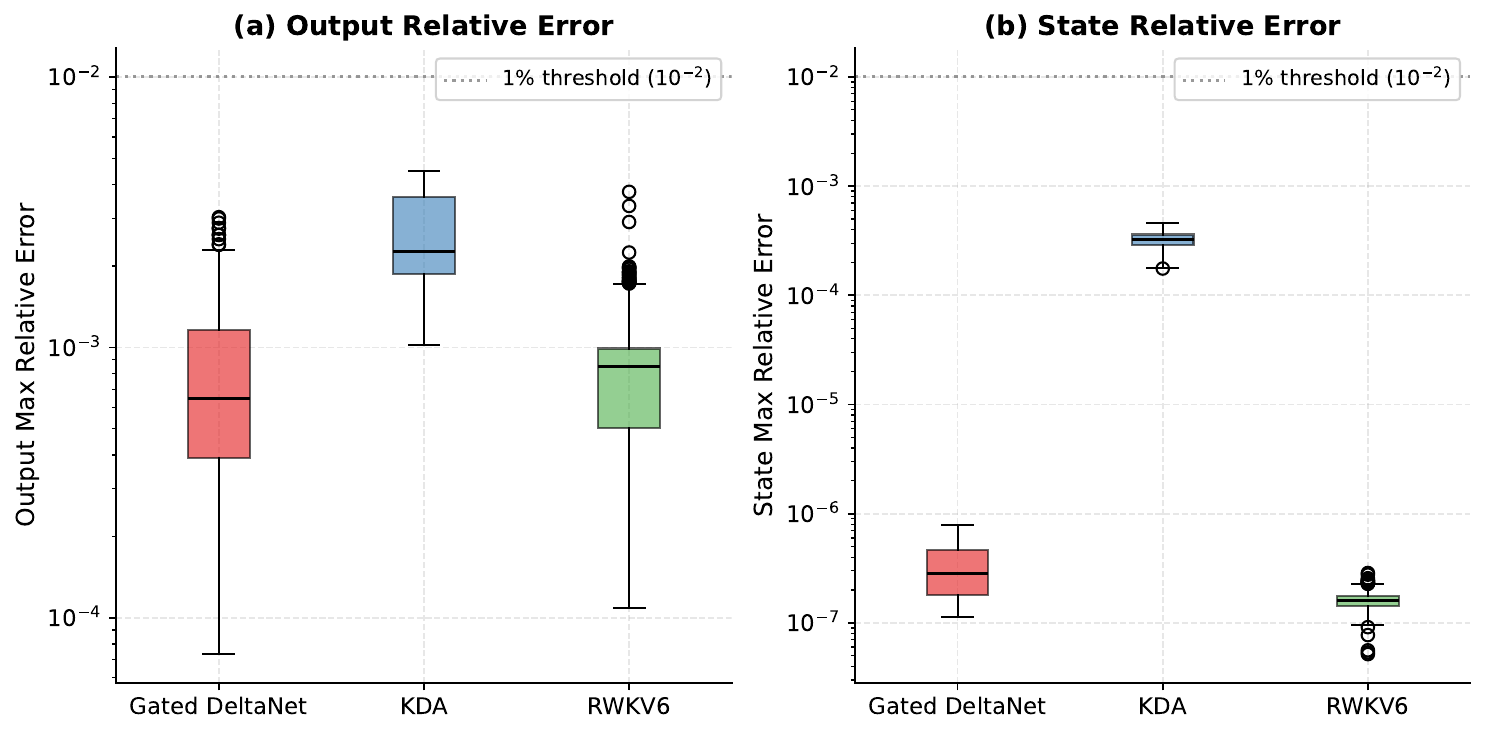}
    \caption{Kernel correctness. (a) Output relative error. (b) Materialized recurrent-state relative error.}
    \label{fig:correctness_error}
\end{figure}

\begin{table}[htbp]
    \centering
    \caption{End-to-end decode TPOT (ms/token) on H200 NVL.}
    \label{tab:e2e_tpot_summary}
    \vspace{-4pt}
    \resizebox{0.95\linewidth}{!}{
    \renewcommand{\arraystretch}{0.94}
    \begin{tabular}{llccc}
        \toprule
        \multirow{2}{*}{\textbf{Model (Operator)}} & \multirow{2}{*}{\textbf{Method}} & \multicolumn{3}{c}{\textbf{Batch Size}} \\
        \cmidrule(lr){3-5}
        & & \textbf{64} & \textbf{128} & \textbf{256} \\
        \midrule
        \multirow{3}{*}{Qwen-35B (GDN)}
         & Baseline & 8.36 & 10.30 & 16.07 \\
         & \method{} & 7.75 & 9.01 & 13.40 \\
         & \textit{Speedup} & \textit{1.08$\times$} & \textit{1.14$\times$} & \textit{1.20$\times$} \\
        \midrule
        \multirow{3}{*}{Kimi-48B (KDA)}
         & Baseline & 7.66 & 8.92 & 13.98 \\
         & \method{} & 7.33 & 8.13 & 12.54 \\
         & \textit{Speedup} & \textit{1.05$\times$} & \textit{1.10$\times$} & \textit{1.12$\times$} \\
        \bottomrule
    \end{tabular}
    }
\end{table}

\subsection{End-to-End Serving Impact}
\label{sec:e2e_serving}

Finally, we evaluate whether the kernel-level gains translate into full-model serving improvements. We integrate \method{} into the vLLM-based serving path~\cite{kwon2023efficient} and measure decode TPOT (ms/token) on H200 for Qwen3.6-35B-A3B and Kimi-Linear-48B-A3B-Instruct. Each run generates 128 decode tokens. The dense baseline uses the same model, batch size, precision, and CUDA-graph replay configuration.

Table~\ref{tab:e2e_tpot_summary} reports TPOT for $B \in \{64,128,256\}$. \method{} improves all six recorded model--batch configurations. On Qwen/GDN at $M=8$, TPOT speedups range from $1.08$--$1.20\times$. On Kimi/KDA at $M=4$, speedups range from $1.05$--$1.12\times$.
The end-to-end gains are smaller than the isolated kernel gains, as expected. Full-model decode includes feed-forward layers, normalization, scheduling, memory management, graph replay, and other framework overheads. \method{} optimizes only the recurrent-state update path, so its end-to-end benefit is bounded by the fraction of TPOT spent in that path, consistent with Amdahl's law~\cite{amdahl1967validity}.

\section{Discussion and Limitations}
\label{sec:discussion_limitations}

\method{} is designed to optimize large-batch recurrent decoding, where recurrent-state write-back can constitute a substantial portion of memory traffic. This focus is reflected in the state-size sweep and end-to-end serving results, and is further corroborated by the H200 GDN counter profile, which confirms the expected reduction in write traffic under a representative configuration.
At the serving level, however, the realized speedups are limited by components outside the optimized update path. In full-model decoding, latency also includes MLP computation, normalization, scheduling, memory management, CUDA-graph replay, and other runtime overheads. Consequently, the end-to-end improvement is bounded by the fraction of execution time affected by \method{}, explaining the gap between isolated kernel speedups and full-model serving gains.
The benefits of \method{} are therefore most pronounced in autoregressive decoding regimes where recurrent-state write-back is a dominant cost. In contrast, workloads with small batches or small recurrent states offer less opportunity for this optimization. Our current serving evaluation focuses on the GDN and KDA paths under a graph-mode runtime, providing initial validation in representative recurrent-decoding settings. Future evaluation across additional operators, deployment configurations, and hardware-counter profiles would further characterize the generality of the method.

\section{Related Work}

\method{} relates to recurrent linear attention, efficient recurrent kernels, and IO-aware inference. In contrast to prior work that modifies architectures or optimizes parallel execution, \method{} reduces decode-time recurrent-state write-back while preserving the trained model.

\textbf{Recurrent linear attention.}
Linear attention~\cite{katharopoulos2020transformers,choromanski2021performer} replaces token-level histories with constant-size recurrent statistics, which fast-weight linear Transformers~\cite{schlag2021linear} interpret as associative memories updated by outer products. Subsequent models improve these recurrences through decay or gating, as in RetNet~\cite{retnet} and GLA~\cite{yang2024gla}; delta-rule corrections, as in DeltaNet~\cite{yang2024parallelizing} and Gated DeltaNet~\cite{gated_deltanet}; and dynamic, matrix-valued, or finer-grained updates, as in RWKV6~\cite{rwkv6} and Kimi Linear~\cite{kimi_linear}. Related fixed-state or subquadratic sequence mixers include state-space and hybrid models such as S4~\cite{gu2022efficiently}, Mamba~\cite{mamba}, Mamba-2~\cite{dao2024transformers}, Jamba~\cite{lieber2024jamba}, and gated convolutional models~\cite{arora2024simple}. Unlike these architecture-level advances, \method{} preserves the recurrence and optimizes its decode-time state representation.

\textbf{Efficient recurrent algorithms and kernels.}
Although recurrent formulations enable efficient one-token decoding, their sequential dependencies complicate parallel training and prefill. Prior work addresses this challenge with parallel or chunkwise formulations of recurrent linear attention~\cite{katharopoulos2020transformers,yang2024gla,yang2024parallelizing,dao2024transformers,qin2024lightning}. Complementary systems efforts, including FLA~\cite{yang2024fla}, ThunderKittens~\cite{spector2024thunderkittens}, FlashRNN~\cite{poppel2024flashrnn}, and MetaAttention~\cite{chen2026metaattention}, provide optimized kernels, reusable abstractions, or code generation for recurrent and linear-attention operators. These methods primarily improve parallel or chunkwise execution with eagerly materialized states, whereas \method{} targets one-token decoding and amortizes dense recurrent-state write-back across steps.

\textbf{IO-aware serving.}
FlashAttention~\cite{dao2022flashattention,dao2024flashattention2,shah2024flashattention3} reduces HBM traffic in softmax attention via IO-aware tiling, and FlashInfer~\cite{ye2025flashinfer} extends this principle to inference kernels and scheduling. Serving systems further improve utilization through batching, KV-cache paging, offloading, chunked prefill, and prefill--decode disaggregation~\cite{yu2022orca,kwon2023efficient,sheng2023flexgen,agrawal2024sarathiserve,zhong2024distserve}. However, these methods primarily target Transformer KV-cache bottlenecks, whereas recurrent linear attention is dominated at decode time by fixed-state write-back. A complementary FPGA design~\cite{gupta2026persistent} avoids this cost by keeping the GDN state on chip for batch-one inference. In contrast, \method{} targets GPU execution, reducing recurrent-state write-back through a base-plus-update-log representation and value-tiled kernels.

\section{Conclusion}

We presented \method{}, a deferred materialization strategy for recurrent linear-attention decoding. By representing the recurrent state as a dense base plus a bounded log of compact updates, \method{} avoids most full dense-state write-backs while preserving the original recurrence semantics. Across GDN, KDA, and RWKV6, \method{} accelerates recurrent-state update kernels by up to 1.86× and reduces profiled recurrent-state write traffic by up to 7.83×. When integrated into a prototype serving stack, \method{} further achieves 1.05–1.20× end-to-end serving speedups over dense recurrent baselines. These results show that recurrent-state materialization is a significant bottleneck in linear-attention decoding, and that optimizing the physical state-update schedule is an effective direction for improving inference efficiency.

\bibliographystyle{plain}
\bibliography{references}

\end{document}